\documentclass[12pt]{iopart}
\usepackage{iopams}
\usepackage{epsfig,amssymb,euscript}

\newcommand{\nn}{\nonumber \\}

\newcommand{\bC}{\mathbb{C}}

\def\te{{\tilde{\epsilon}}}

\begin{document}

%%%%%%%%%%%%%%%%%%%%%%%%%%%%%%%%%%%%%%%%%%%%%%%%%%%%%%%%%%%%%%%%%%%%%%%%%%

%%%%%%%%%%%%%%%%%%%%%%%%%%%%%%%%%%%%%%%%%%%%%%%%%%%%%%%%%%%%%%%%%%%%%%%%%%

\title[Vanishing Preons in the Fifth Dimension]{Vanishing Preons in the Fifth
Dimension}

\author{Jai Grover${}^{1}$, Jan. B. Gutowski${}^{2}$ and Wafic Sabra${}^{3}$}

\address{${}^{1,2}$ DAMTP, Centre for Mathematical Sciences, University of
Cambridge, Wilberforce Road, Cambridge CB3 0WA, UK}
\address{${}^3$ Centre for Advanced Mathematical Sciences and
Physics Department, American University of Beirut, Lebanon}

\ead{${}^1$jg372@cam.ac.uk, ${}^2$J.B.Gutowski@damtp.cam.ac.uk,
${}^3$ws00@aub.edu.lb}

\begin{abstract}

We examine supersymmetric solutions of $N=2$, $D=5$ gauged
supergravity coupled to an arbitrary number of abelian vector
multiplets using the spinorial geometry method.
By making use of methods developed in \cite{papadgran2006b}
 to analyse preons in type
IIB supergravity, we show that there are no solutions preserving
exactly 3/4 of the supersymmetry.

\end{abstract}

%Uncomment for PACS numbers title message
%\pacs{04.50.+h, 04.65.+e, 11.30.Pb}
% Keywords required only for MST, PB, PMB, PM, JOA, JOB?
%\vspace{2pc}
%\noindent{\it Keywords}: Article preparation, IOP journals
% Uncomment for Submitted to journal title message
%\submitto{\JPA}
% Comment out if separate title page not required
%\maketitle

\section{Introduction}

Considerable research activity has been devoted recently to the analysis
and the study of black holes and other gravitational configurations in $N=2$,
$D=5$ gauged supergravity coupled to abelian vector multiplets
\cite{gunaydin:85}. It can be said that this, to a large extent, has been
motivated by the AdS/CFT conjectured equivalence \cite{maldacena}. For
example, string solutions preserving $1/4$ of supersymmetry have been found
in \cite{sabrklemm2003}. Examples of $1/2$ supersymmetric solutions are the
domain wall solutions in \cite{sabrklemm2003}, as well as the solutions
given in \cite{sabrchamsed1998}, \cite{sabrklemm2000a}, \cite{sabrklemm2000b}
and \cite{sabrliu2004} which correspond to black holes without regular
horizons, \ \ \ i. e.,  the solutions either have naked singularities or
closed timelike curves.

More recently, motivated by the method of \cite{tod}, a systematic approach
has been employed in order to classify  $1/4$ supersymmetric
solutions of the minimal gauged five dimensional supergravity
\cite{gutgaunt2003}. The basic idea is to assume the existence of a Killing
spinor, (i.e., to assume that the solution preserves at least one
supersymmetry) and construct differential forms as bilinears in the Killing
spinor. The algebraic and differential conditions satisfied by these forms
are sufficient to determine the local form of the space-time metric
and the rest of the bosonic fields of the theory. This general framework
provides a more powerful method for obtaining many new interesting black
holes than the method of guessing an Ansatz.
The first examples of explicit $1/4$ supersymmetric
regular asymptotically $AdS_{5}$ supersymmetric solutions were given
in \cite{gutreall2004a}. The classification of $1/4$ supersymmetric
solutions and more explicit regular solutions of the gauged supergravity
with vector multiplets were later
given in \cite{gutreall2004b, gutsabra2005}.
Further solutions were considered in \cite{popelu2005} and
\cite{reallluc2006}.

The results obtained in the literature so far seem to have focused mainly on
the classification of supersymmetric solutions of $N=2$, $D=5$ gauged
supergravity which preserve $2$ of the 8 supersymmetries. In $N=2$, $D=5$
gauged supergravity, it is known that the
only solution which preserves all 8 of the supersymmetries is $AdS_{5}$ with
vanishing gauge field strengths and constant scalars.
Moreover, the Killing spinor equations are linear over $\bC$ when
written in terms of Dirac spinors. Hence it follows that
supersymmetric solutions of this theory preserve
either 2, 4, 6 or 8 of the supersymmetries.

In particular, this immediately excludes the possibility of solutions
preserving exactly $7/8$ of the supersymmetry. Such solutions
would be lower-dimensional analogues of hypothetical
preon solutions in $D=11$ supergravity \cite{bandos2001}, which, if possible,
preserve $31/32$ of the supersymmetry.
Properties of preons in ten and eleven dimensions
have also been investigated in
\cite{papadgran2006b, duff2003, bandos2003, bandos2006, hull2003}.
It has also been shown that there are no exactly
3/4 supersymmetric solutions of minimal $N=2$, $D=4$ gauged
supergravity for which one of the Killing spinors generates a null
Killing vector \cite{klemmcald:2004}.

Having eliminated the possibility of preonic solutions
of $N=2$, $D=5$ gauged supergravity, it is natural to
investigate whether solutions preserving the
next highest proportion of supersymmetry,
i.e. exactly $3/4$ supersymmetric solutions, can exist.
In this paper, we present a proof that such solutions also
do not exist.
In order to construct the non-existence proof,
it will be particularly useful to consider the spinors as
differential forms \cite{lawson, wang, harvey}. This method of writing
spinors as forms has been used to classify solutions of supergravity
theories in ten and eleven dimensions (see for example
\cite{papadgran2006b, papadgran2005a, papadgran2005b, papadgran2006a}.)

The plan of the paper is as follows. In Section 2, we review some of the
properties of five-dimensional gauged supergravity coupled to abelian vector
multiplets. In Section 3, we show how spinors of the theory can be written
as differential forms, and how the $Spin(4,1)$ gauge freedom present in the
theory can be used to reduce a spinor to one of three ``canonical''
forms. We also define a $Spin(4,1)$-invariant non-degenerate
bilinear form $B$ on the space of spinors. In Section 4, we show how
solutions preserving $3/4$ of the supersymmetry can be placed into three
classes according as to the canonical form of the spinor which is orthogonal
(with respect to $B$) to the Killing spinors.
This method of
characterizing supersymmetric solutions by the spinors which are
orthogonal to the Killing spinors was originally developed in
\cite{papadgran2006b} where it was used
to show that there are no preons in type IIB supergravity.
For each class of solutions,
we prove that the algebraic Killing spinor equations constrain the solution
in such a manner that the solution reduces to a solution of the
\textit{minimal} gauged five-dimensional supergravity. Finally,
in Section 5, we
show that for all three possible types of solution, the integrability
conditions of the Killing spinor equations in the minimal five-dimensional
gauged supergravity fix the gauge field strengths to vanish, and constrain
the spacetime geometry to be $AdS_{5}$. However, it is known that $AdS_{5}$
is the unique maximally supersymmetric solution of this theory. It therefore
follows that there can be no \textit{exactly} $3/4$ supersymmetric solutions
of $N=2$, $D=5$ gauged supergravity coupled to arbitrary many
vector multiplets.

\section{ $N=2$, $D=5$ supergravity}

In this section, we review briefly some aspects of the $N=2$, $D=5$ gauged
supergravity with field content consisting of
the graviton, the gravitino, $n$ vector potentials, $n-1$ gauginos and
$n-1$ scalars. The bosonic action of
this theory is~\cite{gunaydin:85}
\begin{eqnarray}
\label{bosact}
S &=& {\frac{1 }{16 \pi G}} \int \big( -{}^5 R + 2 \chi^2 {\mathcal{V}}
-Q_{IJ} F^I \wedge *F^J +Q_{IJ} dX^I \wedge \star dX^J   \nn
&-&{\frac{1 }{6}} C_{IJK} F^I \wedge F^J \wedge A^K \big)
\end{eqnarray}
where $I,J,K$ take values $1, \ldots ,n$ and $F^I=dA^I$.
$\chi$ is a nonzero constant, and $C_{IJK}$ are
constants that are symmetric on $IJK$; we will assume that $Q_{IJ}$ is
invertible, with inverse $Q^{IJ}$. The metric has signature $(+,-,-,-,-)$.

The $X^{I}$ are scalars which are constrained via
\begin{equation}
{\frac{1}{6}}C_{IJK}X^{I}X^{J}X^{K}=1\,.  \label{eqn:conda}
\end{equation}
We may regard the $X^{I}$ as being functions of $n-1$ unconstrained scalars
$\phi ^{a}$. In addition, the coupling $Q_{IJ}$ depends on the scalars via
\begin{equation}
Q_{IJ}={\frac{9}{2}}X_{I}X_{J}-{\frac{1}{2}}C_{IJK}X^{K}
\end{equation}
where
\begin{equation}
X_{I}\equiv {\frac{1}{6}}C_{IJK}X^{J}X^{K}
\end{equation}
so in particular
\begin{equation}
Q_{IJ}X^{J}={\frac{3}{2}}X_{I}\,,\qquad Q_{IJ}\partial _{a}X^{J}=
-{\frac{3}{2}}\partial _{a}X_{I}\,.
\end{equation}
The scalar potential can be written as
\begin{equation}
{\mathcal{V}}=9V_{I}V_{J}(X^{I}X^{J}-{\frac{1}{2}}Q^{IJ})
\end{equation}
where $V_{I}$ are constants.

For a bosonic background to be supersymmetric there must be a spinor
$\epsilon $ for which the supersymmetry variations of the gravitino and the
superpartners of the scalars vanish. We shall investigate the properties of
these spinors in greater detail in the next section. The gravitino Killing
spinor equation is

\begin{equation}
\left[ \nabla _{\mu }+{\frac{1}{8}}\gamma _{\mu }X_{I}F^{I}{}_{\rho \sigma
}\gamma ^{\rho \sigma }-{\frac{3}{4}}X_{I}F^{I}{}_{\mu \rho }\gamma ^{\rho }
\right] \epsilon +{\frac{i\chi }{2}}V_{I}(X^{I}\gamma _{\mu }-3A^{I}{}_{\mu
})\epsilon =0  \label{eqn:grav}
\end{equation}
and the algebraic Killing spinor equations associated with the variation of
the scalar superpartners is

\begin{equation}  \label{eqn:newdil}
F^I{}_{\mu \nu} \gamma^{\mu \nu} \epsilon = [ X^I X_J F^J{}_{\mu \nu}
\gamma^{\mu \nu} +2 \gamma^\mu \nabla_\mu X^I ]\epsilon + 4i\chi (X^I V_J
X^J -{\frac{3 }{2}} Q^{IJ}V_J)\epsilon \ .
\end{equation}

We shall refer to ({\ref{eqn:newdil}}) as the dilatino Killing spinor
equation. We also require that the bosonic background should satisfy the
Einstein, gauge field and scalar field equations obtained from the action
({\ref{bosact}}); however we will not make use of these equations in our
analysis, as it will sufficient to work with the Killing spinor equations
alone.

\section{Spinors in Five Dimensions}

Dirac spinors in five dimensions can be written as complexified forms on
$\mathbb{R}^2$ (this construction is also given in an appendix
of \cite{gill2006}). The space of these spinors will be denoted $\Delta =
\Lambda^* (\mathbb{R}^2) \otimes \mathbb{C}$. A generic spinor $\eta$ can
therefore be written as

\begin{equation}
\eta =\lambda 1+\mu ^{i}e^{i}+\sigma e^{12}
\end{equation}%
where $e^{1}$, $e^{2}$ are 1-forms on $\mathbb{R}^{2}$, and $i=1,2$;
$e^{12}=e^{1}\wedge e^{2}$. $\lambda $, $\mu ^{i}$ and $\sigma $ are complex
functions.

The action of $\gamma $-matrices on these forms is given by

\begin{eqnarray}
\gamma _{i} &=&i(e^{i}\wedge +i_{e^{i}}) \nn
\gamma _{i+2} &=&-e^{i}\wedge +i_{e^{i}}
\end{eqnarray}
for $i=1,2$. $\gamma _{0}$ is defined by

\begin{equation}
\gamma _{0}=\gamma _{1234}
\end{equation}
and satisfies

\begin{equation}
\gamma _{0}1=1,\quad \gamma _{0}e^{12}=e^{12},\quad
\gamma _{0}e^{i}=-e^{i}\ \quad i=1,2 \ .
\end{equation}
The charge conjugation operator $C$ is defined by

\begin{equation}
C1=-e^{12}, \quad Ce^{12}=1, \quad Ce^{i}=-\epsilon _{ij}e^{j}\ \quad i=1,2
\end{equation}
where $\epsilon _{ij}=\epsilon ^{ij}$ is antisymmetric with $\epsilon
_{12}=1 $.

We note the useful identity

\begin{equation}
(\gamma _{M})^{\ast }=-\gamma _{0}C\gamma _{M}\gamma _{0}C \ .
\end{equation}

It will be particularly useful to complexify the gamma-operators via

\begin{eqnarray}
\gamma_{p} &=&{\frac{1 }{\sqrt{2}}} (\gamma_p -i \gamma_{p+2}) = \sqrt{2} i
e^p \wedge \nn
\gamma_{\bar{p}} &=& {\frac{1 }{\sqrt{2}}} (\gamma_p +i \gamma_{p+2}) =
\sqrt{2} i i_{e^p} \ .
\end{eqnarray}

\subsection{Gauge transformations and canonical spinors}

There are two types of gauge transformation which can be used to simplify
the Killing spinors of this theory. First, there are local $U(1)$ gauge
transformations of the type

\begin{equation}
\epsilon \rightarrow e^{i\theta }\epsilon  \label{u1gauge}
\end{equation}
for real functions $\theta$, and there are also local $Spin(4,1)$ gauge
transformations of the form

\begin{equation}
\epsilon \rightarrow e^{{\frac{1}{2}}f^{MN}\gamma _{MN}}\epsilon
\end{equation}
for real functions $f^{MN}$.

Note in particular that ${\frac{1 }{2}}(\gamma_{12}+\gamma_{34})$,
${\frac{1}{2}}(\gamma_{13}-\gamma_{24})$ and
${\frac{1 }{2}}(\gamma_{14}+\gamma_{23})$
generate a $SU(2)$ which leaves $1$ and $e^{12}$ invariant and acts on $e^1$
, $e^2$; whereas ${\frac{1 }{2}}(\gamma_{12}-\gamma_{34})$, ${\frac{1 }{2}}
(\gamma_{13}+\gamma_{24})$ and ${\frac{1 }{2}}(\gamma_{14}-\gamma_{23})$
generate another $SU(2)$ which leaves the $e^i$ invariant but acts on $1$
and $e^{12}$. In addition, $\gamma_{03}$ generates a $SO(1,1)$ which acts
(simultaneously) on $1, e^1$ and $e^2, e^{12}$ , whereas $\gamma_{04}$
generates another $SO(1,1)$ which acts (simultaneously) on $1, e^2$ and
$e^1, e^{12}$.

So, one can always use $Spin(4,1)$ gauge transformations to write a
\textit{single} spinor as

\begin{equation}
\epsilon =f 1  \label{eqn:spin1}
\end{equation}
or

\begin{equation}
\epsilon =f e^{1}  \label{eqn:spin2}
\end{equation}
or

\begin{equation}
\epsilon =f(1+e^{1})  \label{eqn:spin3}
\end{equation}
for some real function $f$.

\subsection{A $Spin(4,1)$ invariant bilinear form on spinors}

In order to analyze the $3/4$ supersymmetric solutions it is necessary to
construct a non-degenerate $Spin(4,1)$ invariant bilinear form on the
space of spinors. We first
define a Hermitian inner product on the space of spinors via

\begin{equation}
\langle
z^{0} 1+z^{1}e^{1}+z^{2}e^{2}+z^{3}e^{12},w^{0} 1+w^{1}e^{1}
+w^{2}e^{2}+w^{3}e^{12}\rangle =
{\bar{z}}^{\alpha }w^{\alpha }
\end{equation}%
summing over $\alpha =0,1,2,3$. However, $\langle ,\rangle $ is not
$Spin(4,1)$ gauge-invariant. We define a bilinear form $B$ by

\begin{equation}
B(\eta ,\epsilon )=\langle C\eta ^{\ast },\epsilon \rangle \ .
\end{equation}%
$B$ satisfies the identities

\begin{eqnarray}
B(\eta ,\epsilon )+B(\epsilon ,\eta ) &=&0   \nn
B(\gamma _{M}\eta ,\epsilon )-B(\eta ,\gamma _{M}\epsilon ) &=&0   \nn
B(\gamma _{MN}\eta ,\epsilon )+B(\eta ,\gamma _{MN}\epsilon ) &=&0
\end{eqnarray}%
for all spinors $\eta ,\epsilon $.

The last of the above constraints implies that $B$ is $Spin(4,1)$ invariant.
Note that $B$ is linear over $\mathbb{C}$ in both arguments. $B$ is also
non-degenerate: if $B(\epsilon,\eta)=0$ for all $\eta$ then $\epsilon=0$.

\section{$3/4$ supersymmetric solutions}

We now proceed to examine solutions preserving six out of the eight allowed
supersymmetries. This implies the existence of three Killing spinors, which
we shall denote by $\epsilon_0, \epsilon_1, \epsilon_2$, which are linearly
independent over $\mathbb{C}$.

Suppose we denote the span (over $\mathbb{C}$) of $\epsilon_0$, $\epsilon_1$
, $\epsilon_2$ by $W$. Any complex three-dimensional subspace of
$\mathbb{C}^4$ can be uniquely specified by its one
(complex) dimensional orthogonal
complement with respect to the standard inner product on $\mathbb{C}^4$. It
follows that one can specify $W$ via its orthogonal complement with respect
to $B$. If the one dimensional $B$-orthogonal subspace to $W$ is spanned by
${\tilde{\epsilon}}$, one has

\begin{eqnarray}
W = W_\te = \{ \psi \in \Delta : B(\psi, {\tilde{\epsilon}})=0 \}
\end{eqnarray}

for some fixed non-vanishing ${\tilde{\epsilon}} \in \Delta$. As $B$ is
$Spin(4,1)$ invariant, it will be most convenient to use a $Spin(4,1)$ gauge
transformation in order to write the spinor ${\tilde{\epsilon}}$ in one of
the three canonical forms; i.e. either ${\tilde{\epsilon}}=1$, or
${\tilde{\epsilon}}=e^1$ or ${\tilde{\epsilon}}=1+e^1$ (up to
an overall scaling
which plays no role in our analysis and can be removed).

If ${\tilde{\epsilon}}=1$ then $W$ is spanned by $\eta_0 = 1$, $\eta_1=e^1$,
$\eta_2 = e^2$. If ${\tilde{\epsilon}}=e^1$ then $W$ is spanned by
$\eta_0=1$, $\eta_1=e^1$, $\eta_2=e^{12}$. If
${\tilde{\epsilon}}=1+e^1$ then $W$ is
spanned by $\eta_0 = 1$, $\eta_1=e^1$, $\eta_2 = -e^2+e^{12}$.

In all cases the Killing spinors $\epsilon_0$, $\epsilon_1$, $\epsilon_2$
are related to the spinors $\eta_A$ for $A=0,1,2$ via

\begin{eqnarray}
\epsilon_A = z_A{}^B \eta_B
\end{eqnarray}

where $z$ is a complex $3\times 3$ matrix such that $\det z\neq 0$.

\subsection{Reduction to Minimal Solutions}

The first stage in the analysis is to show that the dilatino Killing spinor
equations ({\ref{eqn:newdil}}) imply that the $3/4$ supersymmetric solutions
correspond to solutions of the minimal theory. In particular, we shall show
that the scalars $X_I$ must be constant, that there exists a nonzero real
constant $\xi$ such that
\begin{eqnarray}
X_I = \xi V_I
\end{eqnarray}
and that the 2-form field strengths $F^I$ satisfy
\begin{eqnarray}
F^I = X^I H
\end{eqnarray}
where $H$ is a closed 2-form.

To show this we first note that the algebraic
constraints ({\ref{eqn:newdil}})
are linear over $\mathbb{C}$. Hence ({\ref{eqn:newdil}}) is equivalent to

\begin{eqnarray}  \label{eqn:newdil2}
F^I{}_{\mu \nu} \gamma^{\mu \nu} \eta_A &=& ( X^I X_J F^J{}_{\mu \nu}
\gamma^{\mu \nu} +2 \gamma^\mu \nabla_\mu X^I )\eta_A
\nn
&+&4i\chi (X^I V_J X^J -{\frac{3 }{2}} Q^{IJ}V_J)\eta_A
\end{eqnarray}

for $A=0,1,2$. In order to compute ({\ref{eqn:newdil2}}), it is first useful
to evaluate ({\ref{eqn:newdil}}) acting on the spinor
$\lambda 1 + \mu^p e^p+ \sigma e^{12}$. We obtain

\begin{eqnarray}  \label{alg1}
- \sqrt{2}i F^I{}_{0m} \mu^m + \sigma F^I{}_{mn} \epsilon^{mn} - \lambda
F^I{}_m{}^m = \nn
X^I (- \sqrt{2}i H_{0m} \mu^m
+\sigma H_{mn} \epsilon^{mn}
-\lambda H_m{}^m)
+ \lambda \partial_0 X^I - \sqrt{2}i \mu^p \partial_p X^I
\nn
+2i \chi (X^I V_J X^J -{\frac{3 }{2}} Q^{IJ} V_J) \lambda
\end{eqnarray}

and

\begin{eqnarray}  \label{alg2}
\sqrt{2} i \sigma F^I{}_{0m} \epsilon^m{}_{\bar{q}} + \sqrt{2}i \lambda
F^I{}_{0 \bar{q}} -F^I{}_m{}^m \mu_{\bar{q}} +2 F^I{}_{m \bar{q}} \mu^m =
 \nn
X^I (\sqrt{2} i \sigma H{}_{0m} \epsilon^m{}_{\bar{q}} + \sqrt{2}i \lambda
H{}_{0 \bar{q}} -H{}_m{}^m \mu_{\bar{q}}
+2 H{}_{m \bar{q}} \mu^m)   \nn
- \partial_0 X^I \mu_{\bar{q}} - \sqrt{2}i \sigma \partial_p X^I
\epsilon^p{}_{\bar{q}} - \sqrt{2}i \lambda \partial_{\bar{q}} X^I + 2i \chi
(X^I V_J X^J -{\frac{3 }{2}} Q^{IJ} V_J) \mu_{\bar{q}}
\end{eqnarray}

and

\begin{eqnarray}  \label{alg3}
- \sqrt{2}i F^I{}_{0 \bar{m}} \epsilon^{\bar{m}}{}_n \mu^n - \lambda
F^I{}_{\bar{m} \bar{n}} \epsilon^{\bar{m} \bar{n}}
+ \sigma F^I{}_m{}^m =
\nn
X^I( - \sqrt{2}i H{}_{0 \bar{m}} \epsilon^{\bar{m}}{}_n \mu^n - \lambda
H{}_{\bar{m} \bar{n}} \epsilon^{\bar{m} \bar{n}}+
\sigma H{}_m{}^m)   \nn
+ \sigma \partial_0 X^I - \sqrt{2}i \partial_{\bar{p}} X^I
\epsilon^{\bar{p}}{}_q \mu^q +2 i \chi (X^I V_J X^J
-{\frac{3 }{2}} Q^{IJ} V_J) \sigma
\nn
\end{eqnarray}

where $\mu_{\bar{q}} \equiv \delta_{\bar{q} p} \mu^p$, and we have defined
$H=X_I F^I$. ({\ref{alg1}}), ({\ref{alg2}}), and ({\ref{alg3}})
correspond to the $1$, $e^q$ and the $e^{12}$ components
of ({\ref{eqn:newdil}}) respectively.

\subsubsection{Solutions with $B$-orthogonal spinors to $1$}

For solutions with spinors $\epsilon_A$ such that $B(\epsilon_A,1)=0$, we
compute the constraints obtained from ({\ref{eqn:newdil2}}), taking
$\eta_0=1 $, $\eta_1=e^1$, $\eta_2=e^2$; using ({\ref{alg1}})-({\ref{alg3}})
to read off the components of the constraints.

Evaluating ({\ref{alg1}}) on $\eta_0=1$ we find the constraint

\begin{eqnarray}
-F^I{}_m{}^m = -X^I H_m{}^m + \partial_0 X^I +2i \chi
(X^I X^J-{\frac{3 }{2}} Q^{IJ}) V_J \ .
\end{eqnarray}

Splitting this expression into its real and imaginary parts we find

\begin{eqnarray}  \label{con1a}
\partial_0 X^I=0
\end{eqnarray}

and

\begin{equation}
\label{newlb1}
F^{I}{}_{m}{}^{m}=X^{I}H_{m}{}^{m}-2i\chi (X^{I}X^{J}-{\frac{3}{2}}
Q^{IJ})V_{J}.
\end{equation}

Evaluating ({\ref{alg1}}) on $\eta_m = e^m$ we find

\begin{eqnarray}  \label{con1c}
F^I_{0m} = X^I H_{0m}+\partial_m X^I \ .
\end{eqnarray}

Next, we evaluate ({\ref{alg3}}) acting on $\eta_0=1$, to
find the constraint

\begin{eqnarray}  \label{con1d}
F^I{}_{mn}= X^I H_{mn}
\end{eqnarray}

and the constraint from ({\ref{alg3}}) acting on $\eta_m=e^m$ is equivalent
to ({\ref{con1c}}).

Finally, we evaluate ({\ref{alg2}}) acting on $\eta_0=1$ to find

\begin{eqnarray}  \label{con1e}
F^I_{0m} = X^I H_{0m}-\partial_m X^I
\end{eqnarray}

and evaluating ({\ref{alg2}}) acting on $\eta_q = e^q$ we find

\begin{eqnarray}
  \label{con1f}
-F^{I}{}_{m}{}^{m}\delta _{p\bar{q}}+2F^{I}{}_{p\bar{q}}
&=&X^{I}(-H_{m}{}^{m}\delta _{p\bar{q}}+2H_{p\bar{q}})
\nn
&-&\partial _{0}X^{I}\delta _{p\bar{q}}+2i\chi (X^{I}X^{J}-{\frac{3}{2}}
Q^{IJ})V_{J}\delta _{p\bar{q}} \ .
\end{eqnarray}

First compare ({\ref{con1c}}) with ({\ref{con1e}}), to find
\begin{equation}
\partial _{m}X^{I}=0 \ .
\end{equation}

This, together with ({\ref{con1a}}) implies that the $X_{I}$ are constant.
Substituting back into ({\ref{con1c}}) we find
\begin{equation}
F_{0m}^{I}=X^{I}H_{0m} \ .
\end{equation}

Next take the trace of ({\ref{con1f}}) to obtain the constraint
\begin{equation}
(X^{I}X^{J}-{\frac{3}{2}}Q^{IJ})V_{J}=0 \ .
\end{equation}
This is equivalent to
\begin{equation}
X_{I}V_{J}X^{J}-V_{I}=0 \ .
\end{equation}

Hence, if $V_{J}X^{J}=0$ at any point, then $V_{I}=0$ for all $I$. As we are
interested in solutions of the gauged theory, we discard this case. Hence
there is a non-zero constant $\xi $ such that
\begin{equation}
X_{I}=\xi V_{I} \ .
\end{equation}
Substituting this back into ({\ref{newlb1}}) we obtain
\begin{equation}
F^{I}{}_{m}{}^{m}=X^{I}H_{m}{}^{m} \ .
\end{equation}
Finally, substituting this back into ({\ref{con1f}}) we find
\begin{equation}
F^{I}{}_{p\bar{q}}=X^{I}H_{p\bar{q}} \ .
\end{equation}

Hence we have the identity
\begin{eqnarray}
F^I = X^I H
\end{eqnarray}
which completes the reduction of these solutions to solutions of the minimal
theory.

\subsubsection{Solutions with $B$-orthogonal spinors to $1+e^1$}

For solutions with spinors $\epsilon_A$ such that $B(\epsilon_A,1+e^1)=0$,
we compute the constraints obtained from ({\ref{eqn:newdil2}}), taking
$\eta_0=1$, $\eta_1=e^1$, $\eta_2=e^{12}-e^2$; using
({\ref{alg1}})-({\ref{alg3}}) to read off the components of the constraints.

Evaluating ({\ref{alg1}}) on $\eta_0=1$ we find the constraint

\begin{eqnarray}
-F^I{}_m{}^m = -X^I H_m{}^m + \partial_0 X^I +2i \chi
(X^I X^J-{\frac{3 }{2}} Q^{IJ}) V_J \ .
\end{eqnarray}

Splitting this expression into its real and imaginary parts we find

\begin{eqnarray}  \label{con2a}
\partial_0 X^I=0
\end{eqnarray}

and

\begin{equation}
\label{newlb4}
F^{I}{}_{m}{}^{m}=X^{I}H_{m}{}^{m}-2i\chi (X^{I}X^{J}-{\frac{3}{2}}
Q^{IJ})V_{J} \ .
\end{equation}

Next, evaluate ({\ref{alg1}}) on $\eta_1=e^1$ to find

\begin{eqnarray}  \label{con2c}
F^I{}_{01} = X^I H_{01} + \partial_1 X^I
\end{eqnarray}

and evaluating ({\ref{alg1}}) on $\eta_2=e^{12}-e^2$ gives

\begin{eqnarray}  \label{con2d}
\sqrt{2}i F^I{}_{02}+ F^I{}_{mn} \epsilon^{mn} = \sqrt{2}i X^I H_{02} + X^I
H_{mn} \epsilon^{mn} + \sqrt{2}i \partial_2 X^I \ .
\end{eqnarray}

Evaluating ({\ref{alg2}}) on $\eta_0=1$ we find

\begin{eqnarray}  \label{con2e}
F^I{}_{0p} = X^I H_{0p} - \partial_p X^I
\end{eqnarray}

and on $\eta_1=e^1$ we obtain (simplifying using ({\ref{con2a}}))

\begin{eqnarray}  \label{con2f}
-F^I{}_m{}^m \delta_{1 \bar{q}} +2 F^I{}_{1 \bar{q}} &=& X^I (-H_m{}^m
\delta_{1 \bar{q}}+2 H_{1 \bar{q}})
\nn
&+&2i \chi (X^I X^J -{\frac{3 }{2}}
Q^{IJ}) V_J \delta_{1 \bar{q}}
\end{eqnarray}

and on $\eta_2=e^{12}-e^2$ we obtain

\begin{eqnarray}
\sqrt{2}iF^{I}{}_{0m}\epsilon ^{m}{}_{\bar{q}}+F^{I}{}_{m}{}^{m}
\delta _{2\bar{q}}-2F^{I}{}_{2\bar{q}} &=&X^{I}(\sqrt{2}iH_{0m}
\epsilon ^{m}{}_{\bar{q}
}+H_{m}{}^{m}\delta _{2\bar{q}}-2H_{2\bar{q}})   \cr
&&-2i\chi (X^{I}X^{J}
-{\frac{3}{2}}Q^{IJ})V_{J}\delta _{2\bar{q}}
\cr
&& -\sqrt{2}i\partial _{m}X^{I}
\epsilon ^{m}{}_{\bar{q}} \ .
\end{eqnarray}

This expression can be further simplified using ({\ref{con2e}}) to give

\begin{equation}
F^{I}{}_{m}{}^{m}\delta _{2\bar{q}}-2F^{I}{}_{2\bar{q}}=X^{I}(H_{m}{}^{m}
\delta _{2\bar{q}}-2H_{2\bar{q}})-2i\chi (X^{I}X^{J}-{\frac{3}{2}}
Q^{IJ})V_{J}\delta _{2\bar{q}} \ .
\end{equation}

Combining this expression with ({\ref{con2f}}) we obtain

\begin{equation}
\label{newlb2}
F^{I}{}_{m}{}^{m}\delta _{p\bar{q}}-2F^{I}{}_{p\bar{q}}=X^{I}(H_{m}{}^{m}
\delta _{p\bar{q}}-2H_{p\bar{q}})-2i\chi (X^{I}X^{J}-{\frac{3}{2}}
Q^{IJ})V_{J}\delta _{p\bar{q}} \ .
\end{equation}

Taking the trace we find that

\begin{equation}
\label{newlb5}
(X^{I}X^{J}-{\frac{3}{2}}Q^{IJ})V_{J}=0 \ .
\end{equation}

Next consider ({\ref{alg3}}) acting on $\eta_0=1$; we obtain

\begin{eqnarray}  \label{con2j}
F^I{}_{mn} = X^I H_{mn}
\end{eqnarray}

and ({\ref{alg3}}) acting on $\eta_1=e^1$ implies

\begin{equation}
F_{02}^{I}=X^{I}H_{02}+\partial _{2}X^{I} \ .
\end{equation}

Combining this with ({\ref{con2c}}) we obtain

\begin{equation}
F_{0p}^{I}=X^{I}H_{0p}+\partial _{p}X^{I} \ .
\end{equation}

However, comparing this expression with ({\ref{con2e}}) it is clear that

\begin{equation}
\partial _{p}X^{I}=0 \ .
\end{equation}

This, together with ({\ref{con2a}}) implies that the $X^{I}$ are again
constant, and then ({\ref{newlb5}}) implies, by the reasoning used in the
previous subsection, that there exists a non-zero constant $\xi $ such that
\begin{equation}
\label{newlb3}
X_{I}=\xi V_{I} \ .
\end{equation}

Then ({\ref{con2e}}) implies
\begin{equation}
F^{I}{}_{0p}=X^{I}H_{0p} \ .
\end{equation}

Next consider ({\ref{newlb2}}). This may be simplified using ({\ref{newlb5}})
to give
\begin{equation}
F^{I}{}_{m}{}^{m}\delta _{p\bar{q}}-2F^{I}{}_{p\bar{q}}=X^{I}(H_{m}{}^{m}
\delta _{p\bar{q}}-2H_{p\bar{q}})
\end{equation}
and then further simplified using ({\ref{newlb4}}) to obtain
\begin{equation}
F_{p\bar{q}}^{I}=X^{I}H_{p\bar{q}} \ .
\end{equation}

Hence it follows that
\begin{eqnarray}
F^I = X^I H
\end{eqnarray}
which completes the reduction of these solutions to those of the minimal
theory.

\subsubsection{Solutions with $B$-orthogonal spinors to $e^1$}

For solutions with spinors $\epsilon_A$ such that $B(\epsilon_A,e^1)=0$, we
compute the constraints obtained from ({\ref{eqn:newdil2}}), taking
$\eta_0=1 $, $\eta_1=e^1$, $\eta_2=e^{12}$; using
({\ref{alg1}})-({\ref{alg3}}) to read off the components of the constraints.

Evaluating ({\ref{alg1}}) on $\eta_0=1$ we find the constraint

\begin{eqnarray}
-F^I{}_m{}^m = -X^I H_m{}^m + \partial_0 X^I +2i \chi
(X^I X^J-{\frac{3 }{2}} Q^{IJ}) V_J \ .
\end{eqnarray}

Splitting this expression into its real and imaginary parts we find

\begin{eqnarray}  \label{con3a}
\partial_0 X^I=0
\end{eqnarray}

and

\begin{equation}
\label{newlb7}
F^{I}{}_{m}{}^{m}=X^{I}H_{m}{}^{m}-2i\chi (X^{I}X^{J}-{\frac{3}{2}}
Q^{IJ})V_{J}.
\end{equation}

Evaluating ({\ref{alg1}}) on $\eta_1=e^1$ and $\eta_2=e^{12}$ we find

\begin{eqnarray}  \label{con3c}
F^I{}_{01} = X^I H_{01} + \partial_1 X^I
\end{eqnarray}
and
\begin{eqnarray}  \label{con3d}
F^I{}_{mn} = X^I H_{mn}
\end{eqnarray}
respectively.

Next consider ({\ref{alg2}}) acting on $\eta_0=1$. This implies

\begin{eqnarray}  \label{con3e}
F^I{}_{0p} = X^I H_{0p} - \partial_p X^I \ .
\end{eqnarray}

({\ref{alg2}}) acting on $\eta _{1}=e^{1}$ together with ({\ref{con3a}})
imply that

\begin{eqnarray}
\label{con3f}
-F^I{}_m{}^m \delta_{1 \bar{q}} +2 F^I{}_{1 \bar{q}} &=& X^I (-H_m{}^m
\delta_{1 \bar{q}} + 2 H_{1 \bar{q}})
\nn
&+& 2 i \chi (X^I X^J -{\frac{3 }{2}}
Q^{IJ}) V_J \delta_{1 \bar{q}}
\end{eqnarray}
and ({\ref{alg2}}) acting on $\eta_2=e^{12}$ is equivalent
to ({\ref{con3e}}).

Next note that ({\ref{alg3}}) acting on $\eta_0=1$ is
equivalent to ({\ref{con3d}}), and
({\ref{alg3}}) acting on $\eta_2=e^1$ implies

\begin{equation}
F^{I}{}_{02}=X^{I}H_{02}+\partial _{2}X^{I}\ .
\end{equation}

Combining this with ({\ref{con3c}}) we obtain

\begin{equation}
F^{I}{}_{0p}=X^{I}H_{0p}+\partial _{p}X^{I}\ .
\end{equation}
Comparing this expression with ({\ref{con3e}}) yields
\begin{equation}
\partial _{p}X^{I}=0
\end{equation}
which together with ({\ref{con3a}}) implies that the $X^{I}$ are constant.
Substituting this back into ({\ref{con3e}}) implies that
\begin{equation}
F^{I}{}_{0p}=X^{I}H_{0p} \ .
\end{equation}

Finally, consider ({\ref{alg3}}) acting on $\eta _{2}=e^{12}$. This implies
\begin{equation}
\label{newlb8}
F^{I}{}_{m}{}^{m}=X^{I}H_{m}{}^{m}+2i\chi (X^{I}X^{J}-{\frac{3}{2}}
Q^{IJ})V_{J} \ .
\end{equation}
Comparing ({\ref{newlb7}}) with ({\ref{newlb8}}) implies that
\begin{equation}
(X^{I}X^{J}-{\frac{3}{2}}Q^{IJ})V_{J}=0 \ .
\end{equation}
As the $X_{I}$ are constant, this constraint implies, using the reasoning in
the previous sections, that there exists a non-zero constant $\xi $ such
that
\begin{equation}
X_{I}=\xi V_{I}
\end{equation}
and hence
\begin{equation}
\label{newlb10}
F^{I}{}_{m}{}^{m}=X^{I}H_{m}{}^{m} \ .
\end{equation}
Then ({\ref{con3f}}) implies that
\begin{equation}
F_{1\bar{q}}^{I}=X^{I}H_{1\bar{q}} \ .
\end{equation}
This constraint, together with ({\ref{newlb10}}) implies that
\begin{equation}
F_{p\bar{q}}^{I}=X^{I}H_{p\bar{q}} \ .
\end{equation}

Hence we have shown that
\begin{equation}
F^{I}=X^{I}H
\end{equation}
which completes the reduction of these solutions to solutions of the minimal
theory.

\section{$3/4$-supersymmetric solutions of the minimal theory}

Having shown that all $3/4$ supersymmetric solutions correspond to solutions
of the minimal theory, it remains to consider the
gravitino Killing spinor equations
of the minimal theory obtained from ({\ref{eqn:grav}}). We substitute
\begin{eqnarray}
X_I = \xi V_I
\end{eqnarray}
into ({\ref{eqn:grav}}) and define $A$ by
\begin{eqnarray}
A = \xi V_I A^I
\end{eqnarray}
so that $H=dA$, with $F^I=X^I H$. Lastly, it is convenient to define
${\hat{\chi}} = \chi \xi^{-1}$, and then drop the hat.

In order to analyze these solutions, we shall consider the integrability
conditions associated with ({\ref{eqn:grav}}). These can be written as

\begin{eqnarray}
\tilde{R}_{MN} \eta_A &\equiv& \big( {\frac{1 }{2}} (S^2_{MN})_{N_1 N_2}
\gamma^{N_1 N_2} + (S^1_{MN})_L \gamma^L \nn &+& {\frac{i }{2}}
(T^2_{MN})_{N_1 N_2} \gamma^{N_1 N_2} + i (T^1_{MN})_L \gamma^L +
{\frac{3i\chi }{2}} H_{MN} \big) \eta_A =0
\end{eqnarray}

for $A=0,1,2$, where

\begin{eqnarray}
(S^2_{MN})_{N_1 N_2} &=& -{\frac{1 }{2}}R_{MNN_1 N_2} -{\frac{1 }{4}}
\epsilon_{L_1 L_2 N_1 N_2 [M} \nabla_{N]} H^{L_1 L_2} \cr &-&{\frac{1 }{4}}
H_{LM} H^L{}_{[N_1} g_{N_2] N} + {\frac{1 }{4}} H_{LN} H^L{}_{[N_1}
g_{N_2] M} +{\frac{3 }{4}} H_{M [N_1} H_{N_2] N} \cr &+&
\big( {\frac{1 }{8}}H_{L_1 L_2} H^{L_1 L_2}+ \chi^2 \big)
g_{M [N_1} g_{N_2] N}
\end{eqnarray}

\begin{eqnarray}
(S^1_{MN})_L = -{\frac{1 }{2}} \nabla_L H_{MN} +{\frac{1 }{4}} H^{L_1 L_2}
H^{L_3}{}_{[M} \epsilon_{N] L_1 L_2 L_3 L}
\end{eqnarray}

\begin{eqnarray}
(T^2_{MN})_{N_1 N_2} = \chi (H_{M [N_1} g_{N_2] N} - H_{N [N_1}
g_{N_2] M})
\end{eqnarray}

and

\begin{eqnarray}
(T^1_{MN})_L = -{\frac{\chi }{4}} \epsilon_{MNL L_1 L_2} H^{L_1 L_2} \ .
\end{eqnarray}

In all cases, we shall show that the integrability condition
${\tilde{R}}_{MN} \eta_A=0$ for $A=0,1,2$ can be used to
obtain constraints involving
only $T^2$, $T^1$ and $H$. These constraints are sufficient to fix $H=0$,
and so $T^1=T^2=S^1=0$. Furthermore, in all cases, the integrability
conditions then imply that $S^2=0$, or equivalently

\begin{eqnarray}
R_{MN N_1 N_2} = 2 \chi^2 g_{M [N_1} g_{N_2] N} \ .
\end{eqnarray}

This implies that the spacetime geometry is $AdS_5$. However, it is known
that $AdS_5$ is the unique maximally supersymmetric solution of this theory.
Hence there can be no solutions preserving \textit{exactly} $3/4$ of the
supersymmetry.

In the following sections, we present the integrability constraints used to
prove this for all three possible types of $3/4$ supersymmetric solutions,
according as whether the Killing spinors $\epsilon_A$ are orthogonal to $1$,
$1+e^1$ or $e^1$. In what follows it will be convenient to suppress the $MN$
indices in the tensors $S^1, S^2, T^1, T^2$ and $H$, though these will be
re-introduced explicitly in several places.

\subsection{Minimal Solutions with $B$-orthogonal spinors to $1$}

The integrability constraints obtained by requiring that
${\tilde{R}}_{MN} 1=0$ are

\begin{eqnarray}
-(S^2)_m{}^m + (S^1)_0 -i (T^2)_m{}^m +i (T^1)_0 +{\frac{3i \chi }{2}} H
&=&0 \cr i (S^2)_{0 \bar{n}} -i (S^1)_{\bar{n}} -
(T^2)_{0 \bar{n}}+ (T^1)_{\bar{n}} &=&0 \cr
(S^2)_{\bar{m} \bar{n}} \epsilon^{\bar{m} \bar{n}} + i
(T^2)_{\bar{m} \bar{n}} \epsilon^{\bar{m} \bar{n}} &=&0
\end{eqnarray}

and the integrability constraints obtained by requiring that
${\tilde{R}}_{MN} e^p =0$ are

\begin{eqnarray}
-i (S^2)_{0p} -i (S^1)_p + (T^2)_{0p} + (T^1)_p &=&0 \cr -(S^2)_m{}^m
\delta_{p \bar{n}} +2 (S^2)_{p \bar{n}} - (S^1)_0 \delta_{p \bar{n}} \cr -i
(T^2)_m{}^m \delta_{p \bar{n}} +2i (T^2)_{p \bar{n}} -i (T^1)_0 \delta_{p
\bar{n}} +{\frac{3i \chi }{2}}H \delta_{p \bar{n}} &=&0 \cr
-i (S^2)_{0 \bar{q}} -i (S^1)_{\bar{q}} + (T^2)_{0 \bar{q}}
+ (T^1)_{\bar{q}} &=&0 \ .
\end{eqnarray}

From these constraints it is straightforward to show that

\begin{eqnarray}
(T^2)_m{}^m =0
\end{eqnarray}
\begin{eqnarray}
(S^2)_m{}^m = 3i \chi H
\end{eqnarray}
\begin{eqnarray}
(S^1)_0 = 0
\end{eqnarray}
\begin{eqnarray}
(T^1)_0 = {\frac{3 \chi }{2}}H
\end{eqnarray}

and

\begin{eqnarray}
(T^2)_{p \bar{q}} = 0
\end{eqnarray}
\begin{eqnarray}
(S^2)_{p \bar{q}} = {\frac{3i \chi }{2}} H \delta_{p \bar{q}}
\end{eqnarray}

and

\begin{eqnarray}
(S^2)_{0p} = - (S^1)_p = i (T^2)_{0p} = -i (T^1)_p \ .
\end{eqnarray}

To proceed, note that imposing the constraint $(T^1_{MN})_0 = {\frac{3 \chi
}{2}}H_{MN}$ for all possible $M,N$ forces all components of $H$ to vanish.
Hence $H=S^1=T^1=T^2=0$, and by the above constraints it follows
that $S^2=0$ also. This implies that the spacetime geometry is $AdS_5$.

\subsection{Minimal Solutions with $B$-orthogonal spinors to $1+e^1$}

The integrability constraints obtained by requiring that
${\tilde{R}}_{MN} 1=0$ are

\begin{eqnarray}
-(S^2)_m{}^m + (S^1)_0 -i (T^2)_m{}^m +i (T^1)_0 +{\frac{3i \chi }{2}} H
&=&0 \cr i (S^2)_{0 \bar{n}} -i (S^1)_{\bar{n}} - (T^2)_{0 \bar{n}}
+ (T^1)_{\bar{n}} &=&0 \cr (S^2)_{\bar{m} \bar{n}}
\epsilon^{\bar{m} \bar{n}} + i
(T^2)_{\bar{m} \bar{n}} \epsilon^{\bar{m} \bar{n}} &=&0 \ .
\end{eqnarray}

The integrability constraints obtained by requiring that ${\tilde{R}}_{MN}
(e^p - \delta^{\bar{p} 2} e^{12}) =0$ are

\begin{eqnarray}
\label{newlb11}
-\sqrt{2} i (S^2)_{0p} - \delta_{p \bar{2}} (S^2)_{mn} \epsilon^{mn}
- \sqrt{2}i (S^1)_p \nn + \sqrt{2} (T^2)_{0p} -i \delta_{p \bar{2}}
(T^2)_{mn}\epsilon^{mn} + \sqrt{2} (T^1)_p &=& 0
\nn
- \sqrt{2} i \delta_{p \bar{2}}(S^2)_{0q} \epsilon^q{}_{\bar{n}}
- (S^2)_m{}^m \delta_{p \bar{n}} +2(S^2)_{p \bar{n}}
\nn
 -(S^1)_0 \delta_{p \bar{n}}
+ \sqrt{2} i \delta_{p \bar{2}} (S^1)_\ell \epsilon^\ell{}_{\bar{n}}
+ \sqrt{2} \delta_{p \bar{2}}
(T^2)_{0q} \epsilon^q{}_{\bar{n}}
\nn
 -i (T^2)_m{}^m \delta_{p \bar{n}}
+2i(T^2)_{p \bar{n}} -i (T^1)_0 \delta_{p \bar{n}}
\nn
- \sqrt{2}
\delta_{p \bar{2}} (T^1)_\ell \epsilon^\ell{}_{\bar{n}}
+{\frac{3i \chi }{2}} H
\delta_{p \bar{n}} &=&0
\nn
 - \sqrt{2} i (S^2)_{0 \bar{q}} \epsilon^{\bar{q}
}{}_p - \delta_{p \bar{2}} (S^2)_m{}^m - (S^1)_0 \delta_{p \bar{2}}
\nn
- \sqrt{2}i (S^1)_{\bar{q}} \epsilon^{\bar{q}}{}_p  + \sqrt{2}
(T^2)_{0 \bar{q}}
\epsilon^{\bar{q}}{}_p -i \delta_{p \bar{2}} (T^2)_m{}^m
\nn
-i (T^1)_0
\delta_{p \bar{2}} + \sqrt{2} (T^1)_{\bar{q}} \epsilon^{\bar{q}}{}_p
-{\frac{3 i \chi }{2}}H \delta_{p \bar{2}} &=&0 \ . \nn
\end{eqnarray}

{}From these constraints we obtain

\begin{eqnarray}
(S^2)_{mn} = i(T^2)_{mn}
\end{eqnarray}

\begin{eqnarray}
(S^1)_0 = i (T^2)_m{}^m
\end{eqnarray}

\begin{eqnarray}  \label{Sptrace}
(S^2)_m{}^m = i(T^1)_0 + {\frac{3 i \chi }{2}} H
\end{eqnarray}

\begin{eqnarray}
(S^2)_{0p} = -i(T^1)_p - {\frac{1 }{\sqrt{2}}} \delta_p{}^2 \epsilon^{mn}
(T^2)_{mn}
\end{eqnarray}

\begin{eqnarray}
(S^1)_p = -i(T^2)_{0p} - {\frac{1 }{\sqrt{2}}} \delta_p{}^2
\epsilon^{mn}(T^2)_{mn}
\end{eqnarray}

\begin{eqnarray}  \label{Spbarq}
(S^2)_{p\bar{q}} &=& i(T^2)_m{}^m \delta_{p\bar{q}} -
\sqrt{2} \delta_{p \bar{2%
}} \epsilon^{\ell}{}_{\bar{q}} (T^2)_{0\ell} + \sqrt{2}\delta_{p \bar{2}}
\epsilon^{\ell}{}_{\bar{q}}(T^1)_{\ell}
\nn
&+& i(T^1)_0 \delta_{p\bar{q}} -
i(T^2)_{p\bar{q}}
\end{eqnarray}

and from the last constraint in ({\ref{newlb11}}) we find

\begin{eqnarray}
\delta_{p \bar{2}} (T^2)_m{}^m - \sqrt{2}i \epsilon_p{}^{\bar{\ell}}
(T^2)_{0\bar{\ell}} + \epsilon^{\bar{m}\bar{n}}\epsilon_{p2}
(T^2)_{\bar{m}\bar{n}}
\nn +\delta_{p \bar{2}} (T^1)_0 -\sqrt{2}i
\epsilon_p{}^{\bar{\ell}} (T^1)_{\bar{\ell}}
+ {\frac{3 \chi }{2}} \delta_{p \bar{2}} H = 0 \ .
\end{eqnarray}

Choosing $p=2$ in the above constraint allows us to express
$H$ in terms of components of $T$;

\begin{eqnarray}
-{\frac{3 \chi }{2}} H_{MN} = -(T^2{}_{MN})_p{}^p - \sqrt{2}i
(T^2{}_{MN})_{01} + (T^1{}_{MN})_0 - \sqrt{2}i(T^1{}_{MN})_1 \ . \nn
\end{eqnarray}

Evaluating this constraint for all possible choices of $M,N$ forces all
components of $H$ to vanish. Hence $H=S^1=T^1=T^2=0$, and by the above
constraints it follows that $S^2=0$ also. This implies that the spacetime
geometry is once more $AdS_5$.

\subsection{Minimal Solutions with $B$-orthogonal spinors to $e^1$}

The integrability constraints obtained by requiring that
${\tilde{R}}_{MN} 1=0$ are

\begin{eqnarray}
-(S^2)_m{}^m + (S^1)_0 -i (T^2)_m{}^m +i (T^1)_0 +{\frac{3i \chi }{2}} H
&=&0 \cr i (S^2)_{0 \bar{n}} -i (S^1)_{\bar{n}} - (T^2)_{0 \bar{n}}+
(T^1)_{\bar{n}} &=&0 \cr (S^2)_{\bar{m} \bar{n}}
\epsilon^{\bar{m} \bar{n}} + i
(T^2)_{\bar{m} \bar{n}} \epsilon^{\bar{m} \bar{n}} &=&0\ .
\end{eqnarray}

The integrability constraints obtained by requiring that ${\tilde{R}}_{MN}
e^1 =0$ are

\begin{eqnarray}
-i (S^2)_{01} -i (S^1)_1 + (T^2)_{01} + (T^1)_1 &=&0 \cr -(S^2)_m{}^m
\delta_{1 \bar{n}} + 2 (S^2)_{1 \bar{n}} -(S^1)_0 \delta_{1 \bar{n}} \cr -i
(T^2)_m{}^m \delta_{1 \bar{n}} +2i (T^2)_{1 \bar{n}} -i (T^1)_0 \delta_{1
\bar{n}} +{\frac{3i \chi }{2}} H \delta_{1 \bar{n}} &=&0 \cr i
(S^2)_{0 \bar{2}} + i (S^1)_{\bar{2}} - (T^2)_{0 \bar{2}}
- (T^1)_{\bar{2}} &=&0 \ .
\end{eqnarray}

The integrability constraints obtained by requiring that ${\tilde{R}}_{MN}
e^{12} =0$ are

\begin{eqnarray}
(S^2)_{mn} \epsilon^{mn} + i (T^2)_{mn} \epsilon^{mn} &=&0 \cr -i (S^2)_{0q}
+ i (S^1)_q + (T^2)_{0q} - (T^1)_q &=&0 \cr -(S^2)_m{}^m -(S^1)_0 -i
(T^2)_m{}^m -i (T^1)_0 -{\frac{3i \chi }{2}}H &=&0 \ .
\end{eqnarray}

From these constraints we find

\begin{eqnarray}
(S^1)_0 = (T^2)_m{}^m = (S^2)_m{}^m = 0
\end{eqnarray}

and

\begin{eqnarray}
(T^2)_{mn}=(S^2)_{mn} =0
\end{eqnarray}

and

\begin{eqnarray}
i (T^1)_0 = -{\frac{3i \chi }{2}} H
\end{eqnarray}

and

\begin{eqnarray}
(S^2)_{1 \bar{n}}+ i (T^2)_{1 \bar{n}} + {\frac{3i \chi }{2}} H
\delta_{1\bar{n}} =0
\end{eqnarray}

which implies that

\begin{eqnarray}
(T^2)_{1 \bar{1}} =0 \ .
\end{eqnarray}

We also find the constraints

\begin{eqnarray}
(S^2)_{01} = (S^1)_1 = -i (T^2)_{01} = -i (T^1)_1
\end{eqnarray}
and
\begin{eqnarray}
(S^2)_{02} = (S^1)_2 = i (T^2)_{02} = i (T^1)_2 \ .
\end{eqnarray}

Finally note that imposing the constraint
$(T^1_{MN})_0 = -{\frac{3 \chi }{2}}H_{MN}$ for all possible $M,N$
forces all components of $H$ to vanish.
Hence $H=S^1=T^1=T^2=0$, and by the above constraints it
follows that $S^2=0$
also. This implies that the spacetime geometry is again $AdS_5$.

\section{Conclusion}

In conclusion, we have studied configurations preserving $3/4$ of
supersymmetry for the theory of $N=2$ five-dimensional gauged supergravity
coupled to abelian vector multiplets. In our analysis we have employed the
method of writing spinors of the theory as differential forms. By exploiting
the $Spin(4,1)$ gauge freedom, it was shown that solutions preserving $3/4$
of the supersymmetry can be placed into three classes. For each class of
solutions, the algebraic Killing spinor equations, coming from the vanishing
of the dilatino supersymmetric variations, reduce our solutions to that of
the \textit{minimal} gauged five dimensional supergravity. Furthermore,
using the integrability conditions of the Killing spinor equations coming
from the vanishing of the gravitino supersymmetric variations, it was shown
that the gauge field strengths must vanish. This means that the spacetime
geometry is the unique maximally supersymmetric $AdS_{5}$ and therefore
there are no \textit{exactly} $3/4$ supersymmetric solutions of five
dimensional supergravity coupled to arbitrary many vector multiplets.

\medskip

\section*{Acknowledgments}

Jan Gutowski thanks the organizers of the XVth Oporto Meeting on Geometry,
Topology and Physics, during which part of this work was completed. Jai
Grover thanks the Cambridge Commonwealth Trusts for financial support. The
work of Wafic Sabra was supported in part by the National Science Foundation
under grant number PHY-0313416.

\end{document}